\documentclass[preprint,eqsecnum,aps,showpacs]{revtex4}

\usepackage{amssymb}
\usepackage{graphicx}

\newcommand{\be}{\begin{equation}}
\newcommand{\ee}{\end{equation}}
\newcommand{\ba}{\begin{eqnarray}}
\newcommand{\ea}{\end{eqnarray}}
\newcommand{\ban}{\begin{eqnarray*}}
\newcommand{\ean}{\end{eqnarray*}}

\begin{document}

%\preprint{HEP/123-qed}

\title{A Continuum Description of Rarefied Gas Dynamics (II)---
The Propagation of Ultrasound}
\author{Xinzhong Chen}
\affiliation{Astronomy Department, Columbia University, New York, NY 10027}
\author{Hongling Rao}
\affiliation{Microeletronics Sciences Laboratories, Columbia University,
New York, NY 10027}
\author{Edward A. Spiegel}
\affiliation{Astronomy Department, Columbia University, New York, NY 10027 }

\date{\today}

\begin{abstract}

The equations of fluid dynamics developed in
paper I are applied to the study of the propagation of ultrasound
waves.  There is good agreement between the predicted propagation
speed and experimental results for a wide range of Knudsen
numbers.

\end{abstract}


\maketitle

\section{Statement of the Equations}

Modern theoretical studies of the influence of dissipation on the
propagation of sound on the basis of the Navier-Stokes equations
may be said to have begun with the work of Kirchhoff \cite{lam32}.  A
principal aim of that and subsequent studies is to determine how the
propagation speed and the rate of dissipation of the waves depend on their
frequencies.  For this problem, the
predictions from the standard Navier-Stokes equations of fluid dynamics
do not agree well with experiments when the periods of the sound waves
become as short as the mean flight times of the particles of the gas,
that is, when we enter the ultrasound regime.

There are two directions from which to enter that regime.  We can begin
with a gas of freely steaming particles and introduce weak interactions
among them.  In that case, we may with Uhlenbeck \cite{uhl63} ask, ``How
is it possible to impose on the random motion of the molecules the ordered
motion ... which a sound wave represents?''  In the modern language of
dynamical systems theory, this could be seen as a problem of
synchronization in which we witness increasing numbers of particles going
into cooperative motion until all are engulfed.  On the other hand, we may
start from the case of continuum mechanics and attempt to extend the
validity of that description to the case of longer and longer mean free
paths.  It is unlikely that in either case we can successfully
traverse the full range of possible conditions, but we may expect to
encounter an interesting transition between the two regimes.

In this paper, we examine how well the fluid dynamical description
of paper I of this series \cite{che00a} extends into the domain where the
particle mean free paths are comparable to the characteristic macroscopic
length scale of the medium.  In paper I, we derived an extension of the
fluid dynamical equations that we hope may offer an improvement of
this kind and, in the present paper, we study their linear form and the
resultant dispersion relation for sound waves.  In this first section, we
restate the equations given in I before going on to the straightforward
determination of the dispersion relation they imply for the linear theory
of sound waves.

The basic form of the macroscopic equations derived from kinetic theory,
are \cite{cer88,gra58,cha61} \begin{eqnarray}
&  & \partial_t \rho + \nabla\cdot(\rho{\bf u}) = 0 \label{cont}
\\ &  & \rho\left(\partial_t {\bf u} + {\bf u}\cdot\nabla {\bf
u}\right) +\nabla \cdot \mathbb{P} = {\bf 0} \label{mom}
\\ &  & {3\over 2}\rho R\left(\partial_t T + {\bf u}\cdot\nabla T \right)
+ \mathbb{P}:\nabla{\bf u} + \nabla \cdot {\bf Q} = 0 \ ,
\label{heat}
\end{eqnarray}
where $\rho$ is mass density, ${\bf u}$ is the velocity field, $T$
is the temperature, $\mathbb{P}$ is the pressure tensor, ${\bf Q}$
is the heat flux vector and the colon stands for a double dot product.
We have not included an external force.

These equations express Newton's laws of motion for a continuum in
phenomenological theories and they are a formal consequence of
most kinetic theories.  Where approaches to the derivation of
these equations from kinetic theory may differ is in the
expressions for the higher moments, $\mathbb{P}$ and ${\bf Q}$.
The derivations from kinetic theory are important since they
provide formulas for the transport coefficients that appear in the
specific expressions for the pressure tensor and the heat flux.
However, not all treatments of the kinetic theory give the same
explicit formulas for $\mathbb{P}$ and ${\bf Q}$, there being
differences of degree and style of the approximations used. Of
course, when the mean free path of the constituent particles is
sufficiently short compared to all macroscopic lengths in the
problem, there is no real disagreement, since the standard
Navier-Stokes forms work well enough for most purposes.  But when
the macroscopic lengths become short and are comparable to the
mean free paths of the particles, those standard results do not
agree with experiment, as we shall see.  Therefore we must ask
whether there is a continuum approximation that may
provide improved treatments of such problems.

To test whether the expressions for $\mathbb{P}$ and ${\bf Q}$ derived in
paper I from the relaxation model of kinetic theory~\cite{bha54,kog69}
may fulfill this need, we here apply them to study of the propagation of
ultrasound.  In the relaxation model, the relaxation time, $\tau$, may be
taken to be of order of the mean flight time of particles, where the mean
speed is of the order of the speed of sound.  Then, we have
\begin{equation}
\tau = {\alpha\over \rho \sqrt{T}} \ . \label{tau} \end{equation}
where $\alpha$ is a constant that depends on the collision cross section
and the gas constant and we have ignored a possible dependence of the
particle cross-section time on particle speed.

The results in I are based on an expansion in $\tau$, up to first order.
Those expansions led to a pressure tensor, \begin{equation}
\mathbb{P} = \left[p - \mu\left({D \ln T \over Dt} + {2\over 3} \nabla
\cdot {\bf u} \right)\right]\, \mathbb{I} - \mu \, \mathbb{E}
 \label{x39} \end{equation} where
\begin{equation}
p = R\rho T \ , \label{press} \end{equation}
$R$ is the gas constant, \begin{equation}
E^{ij} = {\partial u^i\over \partial x_j } + {\partial u^j\over
\partial x_i } - {2\over 3} \nabla \cdot {\bf u} \, \delta^{ij} \ ,
\label{E} \end{equation}
\begin{equation}
\mu=\tau p \label{visco} \end{equation} is the viscosity and
$D/Dt=\partial_t + {\bf u}\cdot\nabla$. The result for $\mu$,
together with (\ref{tau}), implies Maxwell's conclusion that
viscosity does not depend on density for simple gases. For the
heat flux, we obtained \begin{equation} {\bf Q}= -\eta\nabla T  -
\eta T \nabla\ln p - {5\over 2}\mu {D{\bf u} \over Dt} \label{x40}
\end{equation} where $\eta=\frac{5}{2}\mu R$ is the conductivity.
Both (\ref{x39}) and (\ref{x40}) carry errors of order $\tau^2$
that are not indicated explicitly.

These formulae for $\mathbb{P}$ and ${\bf Q}$ are not expressed explicitly
in terms of the fluid fields.  Rather, their expressions involve some of
the same time derivatives of these fields that appear in the fluid
equations.  Here is the central difference between our results and those
obtained in the Chapman-Enskog approach.  In the latter, partial
derivatives with respect to time are eliminated by the use of lower order
results.  Though we do not use that elimination procedure in deriving
the closure relations, we can nevertheless readily recover the
Navier-Stokes results from ours, when $\tau \rightarrow 0$, as we
described in I.  However, even though both theories formally have
first-order accuracy in $\tau$, the results from them are significantly
different at Knudsen numbers of order unity, as we shall see in what
follows.

\section{The Linear Theory}

\ \ \ \ \  We consider the evolution of perturbations on a uniform medium
and define perturbation variables $\phi, \theta$ and
$\varpi$ through the relations \begin{equation}
\rho = \rho_0 (1 + \phi) \ , \qquad T = T_0(1 + \theta) \qquad {\rm and}
\qquad p = p_0(1 + \varpi) \ , \label{1.1} \end{equation}
where $\rho_0$, $T_0$ and $p_0$ are the constant background values of the
thermodynamic fields.  The perturbation quantities all have small
amplitudes with, for example, $|\phi|<<1$.  From the linearization of
(\ref{press}) we obtain \begin{equation} \varpi = \phi + \theta \ .
\label{p1} \end{equation}
We further assume that there is no background motion so that ${\bf u}$ is
small and needs no subscripts.

In analogy with (\ref{1.1}) we write for the pressure tensor, the
linearized form
\begin{equation} \mathbb{P} = p_0 \mathbb{I}+ \mathbb{T}
\end{equation}
where \begin{equation}
\mathbb{T} = \varpi \mathbb{I} - \tau \left(
\partial_t {\theta} + {2\over 3} \nabla\cdot {\bf u}\right)
\mathbb{I} - \tau \mathbb{E} \label{IP1} \end{equation}
where $\tau$ is evaluated for the state variables of the
background medium and $\mathbb{E}$ is small.  Similarly, since
there is no zeroth order heat flux, we get for the linearized
heat flux, \begin{equation}
{\bf Q} = - \eta_0 T_0 \nabla (\theta + \varpi) - {5\over 2} \mu_0
\partial_t {\bf u}.
\label{Q1} \end{equation}
The quantities $\mu_0=\tau p_0$ and $\eta_0$ are the viscosity and
conductivity evaluated in terms of the state variables for the constant
background medium.

The linearization of (\ref{cont}) is \begin{equation}
\partial_t \phi + \nabla \cdot {\bf u} = 0 \ .  \label{lcont}
\end{equation}
A compact form of the linearized (\ref{mom}) is \begin{equation}
\rho_0 \, \partial_t {\bf u} + p_0 \nabla\cdot \mathbb{T} = {\bf 0}.
\label{lmom} \end{equation}
When we take the divergence of (\ref{lmom}) and use (\ref{lcont}),
find that \begin{equation}
\rho_0 \partial_t^2 \phi - p_0 \nabla \nabla : \mathbb{T}= 0 \ .
\label{2mom} \end{equation}
From (\ref{IP1}) we find that \begin{equation}
\nabla\cdot \mathbb{T}  = \nabla \varpi - \tau \partial_t
\nabla \theta - {2\over 3} \tau \nabla(\nabla\cdot {\bf u)}
- \tau \nabla\cdot \mathbb{E} \ . \label{divP} \end{equation}
We see from (\ref{E}) that \begin{equation}
\nabla \cdot \mathbb{E} = \nabla^2 {\bf u} + {1\over 3}
\nabla(\nabla \cdot {\bf u}). \label{divE} \end{equation}
Then, with the help of (\ref{lcont}), we find \begin{equation}
\nabla\cdot \mathbb{T}  =  \nabla \varpi - \tau \partial_t
\nabla (\theta - \phi) - \tau \nabla^2 {\bf u} \ . \label{dull}
\end{equation}

On using (\ref{lcont}) again, we find that
\begin{equation}
\nabla \nabla : \mathbb{T} =  \nabla^2 \varpi - \tau \partial_t
\nabla^2 (\theta-2\phi) \ , \label{nabla2} \end{equation}
which we may introduce into (\ref{2mom}).  Next we define the Laplacian
speed of sound, $a$, and the kinematic viscosity, $\nu_0$, as in
\begin{equation}
a^2 = {p_0\over \rho_0} \qquad {\rm and} \qquad \nu_0 = {\mu_0\over
\rho_0} \ . \label{defs} \end{equation}
We then obtain the dissipative wave equation \begin{equation}
\left(\partial_t^2 - a^2 \nabla^2 - 2 \nu_0 \partial_t \nabla^2\right)
\phi + \left(\nu_0\partial_t - a^2\right)\nabla^2 \theta = 0 \ .
\label{wave} \end{equation}

To complete this discussion, it is useful to introduce the thermal
diffusivity \begin{equation} \kappa_0 = {\eta_0 \over \rho_0 C_p} \ ,
\label{kappa} \end{equation} where $C_p={5\over 2} R$. Thus, $\eta_0
T_0={5\over 2}p_0\kappa_0$. This is used in the linearized heat
equation, \begin{equation}
{3\over 2} p_0 \partial_t \theta + p_0 \nabla\cdot {\bf u} +
\nabla\cdot {\bf Q} = 0, \label{lheat} \end{equation}
where we may write \begin{equation}
\nabla \cdot {\bf Q} = - {5\over 2} p_0 \kappa_0 \nabla^2(\phi + 2\theta)
+{5\over 2} \rho_0 \nu_0 \partial_t^2 \phi \ . \label{divQ} \end{equation}
Since $\nu_0 = a^2 \tau$, we then find \begin{equation}
\left(\tau \partial_t - {10\over 3}{a^2 \tau^2\over \sigma}
\nabla^2\right) \theta + \left({5\over 3} \tau^2 \partial_t^2
-{2\over 3} \tau\partial_t - {5\over 3} {a^2\tau^2\over \sigma} \nabla^2
\right) \phi = 0  \label{hot} \end{equation}
where \begin{equation}
\sigma = {\nu_0\over \kappa_0}  \label{prandtl} \end{equation}
is the Prandtl number of the undisturbed medium.

Finally, to further simplify the appearance of these formulae, we let
the unit of time be $\tau$ and the unit of length be $a\tau$.  Then our
linearized equations for sound waves are
\begin{equation}
\left(\partial_t^2 - \nabla^2 - 2 \partial_t\nabla^2\right)\phi
+\left(\partial_t - 1\right)\nabla^2 \theta = 0 \ , \label{1}
\end{equation}
\begin{equation}
\left(5\partial_t^2 - {5\over \sigma}\nabla^2 - {2} \partial_t\right)
\phi + \left(3 \partial_t - {10\over \sigma}\nabla^2\right)\theta = 0
\ . \label{2} \end{equation}
For comparison we note that the analogous linear equations for the
Navier-Stokes case (with zero bulk viscosity) are these:
\begin{equation}
\left(\partial_t^2 - \nabla^2 - {4\over 3}
\partial_t\nabla^2\right)\phi - \nabla^2 \theta = 0 \ , \label{Xa}
\end{equation}
\begin{equation}
- {2} \partial_t \phi + \left(3 \partial_t - {5\over \sigma}
\nabla^2\right)\theta = 0
\ . \label{Ya}
\end{equation}

\section{The Dispersion Relations}

We may seek solutions to the linear equations (\ref{1})-(\ref{2})
in which $\phi$ and $\theta$ vary like $\exp\left(ikx- s
t\right)$. Since the mean free path is the unit of length, the
wave number $k$, which is nondimensional, is effectively the
Knudsen number for this problem.  The dispersion relation is
\begin{equation} (3+5k^2) s^3 - \left({10\over \sigma}
-1\right)k^2 s^2 + 5 \left({5\over \sigma} k^2 +1\right)k^2 s -
{5\over \sigma} k^4 = 0 \ .  \label{disp} \end{equation} For
comparison, we report that the dispersion relation for the
Navier-Stokes equations is \begin{equation} 3 s^3 - \left({5\over
\sigma} +4\right)k^2 s^2 + 5 \left({4\over 3\sigma} k^2
+1\right)k^2 s - {5\over \sigma} k^4 = 0 \ .  \label{dispa}
\end{equation}

\subsection{Frequencies}
To get a feeling for what these results mean, we look at free
modes for which $k$ is real.  Then we set $s=i\omega + \alpha$
where $\alpha$ and $\omega$ are also real.  When we introduce this
into (\ref{disp}) we find that there is a (thermal) mode with
$\omega=0$ and a pair of (acoustic) modes whose frequencies
satisfy
\begin{equation} \omega^2 = 3 \alpha^2 -
{2(10-\sigma)\over\sigma(3+5k^2)} k^2 \alpha +{5(\sigma+5k^2)\over
\sigma(3+5k^2)} k^2 \ , \label{freq} \end{equation} which gives
the frequencies of sound waves.  As we may confirm, $\alpha$ is
of order unity for large $k$ and it grows in proportion to $k^2$
for small $k$.  Hence, for both very large and very small $k$, the
last term on the right of (\ref{freq}) is the largest one on that
side.  So we may write the uniform approximation \begin{equation}
\omega^2 = {5(\sigma+5k^2)\over \sigma(3+5k^2)} k^2 \ .
\label{freaq}
\end{equation}

For small $k$, this gives the phase speed $\omega/k= \pm \sqrt{5/3}$,
which is the usual speed of sound for an adiabatic sound wave, as is to be
expected for very long wave lengths.   For large $k$, we obtain the phase
speed $\omega/k = \pm \sqrt{5/\sigma}$.

For the N-S equations with zero bulk viscosity there is the same number
of modes: a thermal mode with zero frequency and
sound waves with \begin{equation} \omega^2 = 3\alpha^2 -
{2\over 3}({5\over \sigma} + 4) k^2 \alpha
+ {5\over 3} ({4k^2\over 3\sigma}+1) k^2 \ . \label{NSO}
\end{equation}
As expected, the two sets of equations agree in the limit of very small
$k$, where the N-S equations return the phase speed $\pm\sqrt{5/3}$.  But
for large $k$, the differences between the two theories become
qualitative.  With the Navier-Stokes equations, we find that at large $k$,
instead of reaching a finite limit, the phase speed is proportional to
$k$ for large $k$.  As we shall see when we look at the experimental
results, the N-S prediction is qualitatively wrong; the phase speed goes
to a finite value at large $k$.

\subsection{Damping Rates}

The equation for the damping rate is \begin{equation}
(3+5k^2)\alpha^3 - ({10\over \sigma}-1)k^2 \alpha^2 - \left[3
(3+5k^2)\omega^2 - 5(1+{5\over \sigma} k^2)k^2\right]\alpha +
({10\over \sigma}-1)k^2 \omega^2 - {5\over \sigma} k^4 = 0.
\label{damp} \end{equation}

For the thermal mode, for which $\omega=0$, we find the damping rates
$\alpha=k^2/\sigma$ for small $k$ and $\alpha=1/5$ for large $k$.  Thus,
there is very little damping for long waves while short waves are damped
on the collisional time scale.  Moreover, on examination of these two
limits, we see that they each emerge from the balance of the same two
terms in (\ref{damp}).  Hence we may write the approximate formula
\begin{equation}
\alpha = {k^2\over \sigma + 5 k^2} \label{alpha} \end{equation}
as a reasonable approximation to the damping rate for all $k$, in the
thermal mode.

Similarly, in the case of sound waves, we see that $\alpha$ is also the
result of the balance between the same two terms in (\ref{damp}) in the
limits of for large and small $k$.  Hence, we find that for sound waves,
to good approximation, the damping rate is given by \begin{equation}
\alpha = {(10 - \sigma)k^2\omega^2 - 5k^4\over
3\sigma(3+5k^2)\omega^2-5k^2(\sigma+5k^2)} \ , \label{alfa} \end{equation}
where $\omega$ is given in (\ref{freaq}).  For long wave lengths, the
damping is again
slight since it goes to zero like $[(7-\sigma)/(6\sigma)]k^2$.  For large
$k$, we obtain the finite limit $\alpha=(5-\sigma)/(5\sigma)$.

For the Navier-Stokes equations, the damping rate is given by
\begin{equation}
3\alpha^3 - ({5\over \sigma} + 4)k^2 \alpha^2 +
\left[5({4k^2\over 3\sigma}+1)k^2 - 9 \omega^2\right]\alpha +
({5\over \sigma}+4)k^2\omega^2 - {5\over \sigma}k^4 = 0 \ . \label{N-S}
\end{equation}
The damping rate for the acoustic modes also goes like $k^2$ for
small $k$ for both the thermal mode and the acoustic modes.  For
the acoustic modes, the damping rate is $\alpha=[(2\sigma +
1)/(3\sigma)]k^2$, so that we get the same wave number dependence,
but with a different coefficient than is obtained from our
equations in the small $k$ limit.  However, for increasing $k$,
the N-S damping rates {\it grow} like $k^2$ for sound waves, which is
in disagreement with experiment~\cite{bha67}.

\section{Comparison with Experiment}

Though the study of free modes in the previous section is
intuitively clear, it does not directly represent the way
experiments on sound propagation are usually carried out.  In the
experiments, it is more typical that one drives the fluid at a
real, fixed frequency and then studies the propagation of waves in
space. The forcing may be accomplished by vibrating the end wall
of a tube containing gas at a fixed (real) frequency $\omega$ and
observing the propagation down the tube.  To model this procedure
in full detail would involve a careful treatment of the forcing
procedure, which usually requires attention to boundary
conditions.  However, in this first reconnaissance of the way our
equations describe sound waves, we shall adopt a standard
theoretical practice \cite{lam32} and simply fix the wave
frequency, $\omega$, in the dispersion relation and compute the
resulting $k$, which will typically be complex.  Thus, in
(\ref{disp}) we let $s=i\omega$ and we find that the equation
for $k$ becomes
\begin{equation} {5\over \sigma}\left(1 -
5i\omega\right)k^4 + \left[5i\omega^3 - 5i\omega - ({10\over
\sigma} -1) \omega^2\right]k^2 + 3i\omega^3 = 0. \label{spatial}
\end{equation}
We may similarly obtain such an equation for the N-S case,
(\ref{dispa}). In order to emphasize the results for large Knudsen
number we plot the results in the manner used, for example, by
Cercignani~\cite{cer88}. That is, we introduce the quantity
\begin{equation} K^2 = {5k^2\over 3\omega^2} \ ,  \label{K2}
\end{equation} where $K$ is a normalized inverse propagation
speed.  The factor $5/3$ is included so that the phase speed is
nondimensionalized on the Laplacian (or adiabatic) speed of sound,
rather than the Newtonian (or isothermal) speed of sound as above.
Then we find that \begin{equation} \left(1 - 5i\omega\right)K^4 +
{\sigma\over 3} \left[5i\left(\omega - {1\over \omega}\right) -
\left( {10 \over \sigma} - 1 \right) \right]K^2 + {5 \sigma i\over
3\omega} = 0. \label{K4}
\end{equation}

\begin{figure}[hbt]
\begin{center}
\includegraphics[width=6cm,angle=270]{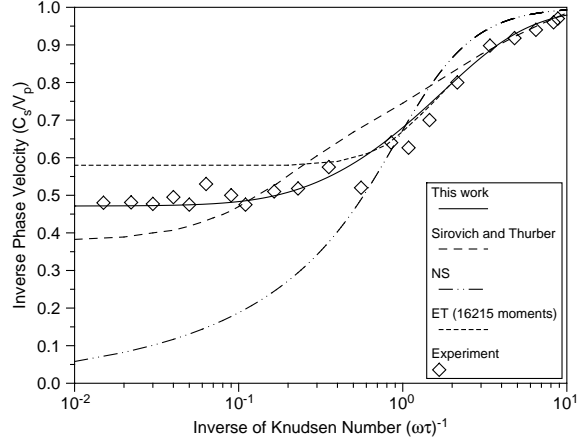}
\end{center}
\caption{The inversed phase velocity as a function of inversed
frequency}
\label{thefig}
\end{figure}

To see how this representation contains the results for free
modes, we note that, in the limit $\omega \rightarrow 0$,
(\ref{K4}) reduces to $K^2 = 1$. That is, for low frequencies, the
usual adiabatic sound speed is recovered.  In the (more
interesting) opposite limit, $\omega \rightarrow \infty$, we
obtain $K^2=\sigma/3$ for the propagative modes.  Thus we see
that, in the limit of forcing at high frequency, sound waves
propagate with phase speeds $\pm \sqrt{\sigma/3}$, independently
of frequency.  The data shown in the accompanying figure (Fig.~\ref{thefig}) confirm
this independence of frequency (or wave number) of the speed of
propagation of ultrasound. The observed nondimensional phase speed
is $0.47$.

The Prandtl number found from the relaxation model of kinetic
theory, either by the methods of Chapman and Enskog or those
described in paper I, is unity. With this value, we obtain $0.51$
for the limiting phase speed, so this represents a small
quantitative error.  However, the value of $\sigma$ found in
kinetic theory depends on the atomic model used, that is, on the
nature of the collision term.  Though the relaxation model gives
the explicit value unity for $\sigma$, the value found with the
traditional Boltzmann equation for hard spheres is $2/3$.  This
difference has nothing to do with the approximation method (our
procedure gives $\sigma=2/3$ when applied to the Boltzmann
equation) but is a consequence of the nature of the form of the
atomic interactions that is adopted.  We therefore follow a common
practice put the empirical Prandtl number into the theoretical
results when comparing with experiments. Since the experimental
data we shall refer to are for noble gases whose values of
$\sigma$ are $0.6$ or $0.7$ we shall here adopt the value
$\sigma=2/3$ suggested by the Boltzmann equation.  When we use that
value of the Prandtl number in evaluating the phase speed, we
obtain $K= \sqrt{2}/3 = 0.471$.  Even without this adjustment, the
results for the propagation of ultrasound are good, but we would
propose to anyone thinking of using our equations from this
first-order development from the relaxation equation to introduce
this phenomenological improvement of the theory.

In the accompanying figure (Fig.~\ref{thefig}),  we show the variation of $\Re e\; K$
as a function of $1/\omega$ from a number of sources.  The
experimental values (\cite{gre56,mey57}) are indicated as
individual points (the diamonds) and they appear to be tending
toward a nonzero constant value at high frequency.  This is
qualitatively in accord with our results, here shown as a solid
line for the case of $\sigma=2/3$, and it is in stark disagreement
with the prediction from the Navier-Stokes equations (long
dashes with double dots), which predict that $\Re e\;K$ goes to zero like
$1/\omega$. Since our limiting value for $\Re e\;K$ was found to
be $\sqrt{\sigma/3}$, the remarkable agreement of our results with
experiment owes something to our using the experimental value of
$2/3$ for $\sigma$ for limiting value $\Re e\; K \rightarrow
0.47$. Nevertheless, even without this choice, the results would
be adequate and comparable to those shown for the moment method
(\cite{mul98}) with $16,215$ moments (short dashes). Other
theoretical studies of ultrasound are based on direct solution of
the Boltzmann equation \cite{uhl63,cer88} and we show the results
of Sirovich and Thurber \cite{sir65} obtained in this way (medium
dashes), for which the Prandtl number automatically has the value
$2/3$.

\section{Conclusion}

\ \ \ \ \ In the study of the thermal damping of sound waves by
electromagnetic radiation~\cite{ste67}, one finds that, for thermal times
much less than the acoustic period, sound propagates at the isothermal
speed of sound with negligible dissipation.  In the opposite limit of long
thermal times, there is also little dissipation, but propagation
is at the adiabatic speed of
sound.   The experiments, and the solution of the Boltzmann
equation show similar behavior when the relevant parameter is the
ratio of the collisional relaxation time to the acoustic period.
Our equations, as well as those of the moment method (with tens of
thousands of moments), reproduce this behavior but the
Navier-Stokes equations do not.  Moreover, when the Prandtl number
is chosen to be that of the experimental gas, the quantitative
agreement becomes very good.  In the next installation of this
series, we shall compute the profile of a stationary shock wave.
As we shall see, the agreement with the experiments is good in
that case too.

\goodbreak

\end{document}